\documentclass[preprint,tightenlines,eqsecnum,floats,aps,amsmath,amssymb,nofootinbib,prd,showpacs]{revtex4}
\usepackage{amsmath,amssymb,amsfonts}
\usepackage{graphicx}
\usepackage{enumerate}
\usepackage[colorlinks=true,linkcolor=blue,anchorcolor=blue,citecolor=blue,urlcolor=blue]{hyperref}
\usepackage[sort&compress]{natbib}
\usepackage{url}
\usepackage[T1]{fontenc}
\usepackage{mathptmx}

\begin{document}
\title{New quasi-universal relations for static and rapid rotating neutron stars}

\author{Wenjie Sun}
\affiliation{School of Physics and Optoelectronics, South China University of
Technology, Guangzhou 510641, P.R. China}
\author{ Dehua Wen\footnote{Corresponding author. wendehua@scut.edu.cn}}
\affiliation{School of Physics and Optoelectronics, South China University of
Technology, Guangzhou 510641, P.R. China}
\author{Jue Wang}
\affiliation{School of Physics and Optoelectronics, South China University of
Technology, Guangzhou 510641, P.R. China}

\date{\today}

\begin{abstract}
In the last few decades, lots of universal relations between different global physical quantities of neutron stars have been proposed to constrain  the unobservable or hard to be observed properties of neutron stars. But  few of them are related to the  gravitational redshift or the gravitational binding energy, especially for the fast rotating neutron stars. Here we will focus on the universal relations related to these two quantities. Based on 11 equations of state (EOSs) from the predictions of microscopic nuclear many-body theories for normal or hybrid neutron stars, we  proposed a set of new quasi-universal relations under three rotating cases: static, general rotating and Keplerian  rotating.  These new quasi-universal relations provide a potential way to constrain  or estimate the  unobservable or hard to be observed  properties of neutron stars.
\end{abstract}


\maketitle
\section{Introduction}
Neutron stars are the densest stars in the universe and their super high density,  ultra high pressure and  extremely strong gravitational field  cannot be reproduced in terrestrial laboratories  at present \cite{1,2,3}, which makes neutron stars the only ideal laboratory to study the fundamental physics under such extreme conditions. For example, according to the information obtained from different observation channels, such as binary star pulse X-rays \cite{4,5} and gravitational waves from a binary neutron star inspiral (GW170817) \cite{7,8},
 people can extract the knowledge of global properties, internal structures and dense matter of neutron stars through the methods such as
   maximum mass constraint \cite{48} and Bayesian analysis \cite{6,Lim,De,Carson,Kastaun2019,Carreau,Lim 2,Lim 3,Carreau 2,Hernandez,Fasano,Riley}.
However, due to the limitation of astronomical observation accuracy, number of observation cases, nuclear physics theory and ground laboratory conditions, there is still great uncertainty in determining some of the properties of the neutron star (such as the radius) and in understanding  the equation of state of the super dense nuclear matter \cite{Lattimer2007,Li2008,5,Feng2018,Li2019EPJA}.

Fortunately,  it has been found that the universal relations between different physical quantities of neutron stars, which are independent of specific equations of state (EOS), provide a way to constrain or estimate the unobservable or hard to be observed properties of neutron stars, and can be used to further investigate the internal properties of neutron stars \cite{9,10,11,Jiang2019,Li2019EPJA}.
During the last decades, lots of attempts have been made to establish precise universal relations.  Many of the earlier pioneering works of the universal relations come from the study of the quasi-normal oscillations \cite{Andersson1996,Andersson1998,Benhar2004,Tsui2005}.  This kind of universal relations builds a bridge to link the parameters of the oscillation modes and some of the global properties of neutron stars, which is essentially independent of the specific EOSs \cite{Andersson1996,Andersson1998,Benhar2004,Tsui2005,Wen2009,16}. Another kind of interesting universal relations relates to the binding energy \cite{17,18,19,Jiang2019}, which can be  traced back to the earlier pioneering work of Lattimer and Yahil \cite{Lattimer1989}. They provide a way to understand some of the internal properties of neutron stars, such as the total binding energy, the nuclear binding energy and the gravitational binding energy. The further  work of the universal relation extends to the rotating stars. For instance, Yagi \& Yunes established a set of interesting and important universal relations, the I-LOVE-Q universal relations, which connected three parameters of the rotation rate (that is, the moment of inertia ($I$), the Love number ($\lambda$) and the quadrupole moment ($Q$)) \cite{9,12,13}. These universal relations can be used to constrain the deformability  through the observed  moment of inertia and to distinguish quark stars from neutron stars  \cite{13}. Recently, Jiang \& Yagi further derived  the I-Love-C relations analytically, which leads to a better understanding of the origin of the universal relations \cite{Jiang2020}.
 There are also some other kind of universal relations for the  rotational neutron stars \cite{20,21,22,23,24,25,26}. For example, Breu \& Rezzolla proposed the universal  relation between the angular momentum  and the  compactness, and found that this  relation is hold at both low and fast spin frequencies \cite{20}.  In Ref. \cite{24}, the universal relations among the gravitational mass, the rest mass and the angular momentum are  established. A set of universal relations of the Keplerian (mass-shedding) rotating neutron stars is present in Ref. \cite{26}.

As stated above, the universal relations between the properties of neutron star   provide valuable clue to  understand the neutron star physics. As far as we know, there are few works focused on the universal relations related to the redshift and the gravitational binding energy, especially for the fast rotating neutron stars \cite{Jiang2019,27}.
In this work, we will  propose some new quasi-universal relations based on these two quantities while considering the spin factor.

The plan of this paper is as follows. In Sec. \ref{sec2}, we briefly review the theoretical framework of static and rotating neutron stars. Then we concisely introduce the adopted EOS models in Sec. \ref{sec3}. The main results of this work are presented in Sec. \ref{sec4} by three parts  (that is, the static, general rotating and Kepler rotating cases). Finally, a brief summary is given  in Sec. \ref{sec5}. In the analytic formula, we use the geometric units (G = c = 1).


\section{Theoretical framework}\label{sec2}

In this section, the theoretical framework and the definition of some global physical quantities of the static and rotating neutron stars are briefly reviewed.

\subsection{Static neutron star}
The equilibrium model of the neutron star under static conditions is a spherically symmetric solution of the Einstein field equation. In this case, the line element is given by
\begin{equation}
d s^{2}=e^{2\nu}dt^{2}-e^{2\lambda}dr^{2}-r^{2}d\theta^{2}-r^{2}\sin^{2}\theta d\phi^{2},
\end{equation}
where $\lambda$ and $\nu$ are functions  depending only on the radius $r$. Normally,  the matter inside a cold neutron star is assumed to  be composed of   perfect fluid,  and thus the energy momentum tensor can be given by
\begin{equation}
T^{\sigma\tau}=p g^{\sigma\tau}+(p+\varepsilon) u^{\sigma} u^{\tau},
\end{equation}
where $\epsilon$, $p$, $u^{\sigma}$ and $g^{\sigma\tau}$ are the energy density, the pressure, the four-velocity and the metric tensor, respectively. According to the Einstein field equation
\begin{equation}
R^{\sigma\tau}-\frac{1}{2} g^{\sigma\tau} R=8 \pi T^{\sigma\tau},
\end{equation}
combining with the  line element of  static spherical symmetric space and the energy momentum tensor of   perfect fluid, we can get the well known TOV equations  \cite{Oppenheimer1939,Tolman1939}
\begin{equation}\label{TOV1}
\frac{\mathrm{d} p}{\mathrm{d} r}=-\frac{(p+\varepsilon)\left[m(r)+4 \pi r^{3} p\right]}{r[r-2 m(r)]},
\end{equation}
and
 \begin{equation} \label{TOV2}
  \frac{{dm(r)}}{{dr}} = 4\pi \varepsilon {r^2},
\end{equation}
where  $m(r)$ is the gravitational mass within  radius $r$. To numerically solve the TOV equations, we should  integrate the Eqs. (\ref{TOV1}) and (\ref{TOV2}) from the center ($m=0$, $r=0$, $\varepsilon = \varepsilon_c$) to the surface ($p=0$, $r=R$ and $m(R)=M_g$, where $M_g$ is the gravitational mass of neutron star).

Through the definition of the proper mass of a neutron star \cite{Cameron1959,Bagchi2011}
\begin{equation}
M_p=\int_{0}^{R} \varepsilon(r)4\pi r^2[1-\frac{2m(r)}{r}]^{-\frac{1}{2}}\, dr,
\end{equation}
we can get the gravitational binding energy as
\begin{equation}
E_g=M_g-M_p.
\end{equation}
Obviously, $E_g<0$.

For a static neutron star, its gravitational redshift is given by
\begin{equation}\label{GZ}
z=\left(1-\frac{2 M_g}{R}\right)^{-1/2}-1.
\end{equation}

\subsection{Rotating neutron star}
The neutron star will be deformed by the rapid rotation. Assuming that the deformed neutron star is still axisymmetric, its line element can be given by \cite{31}
\begin{equation}
\begin{split}
d s^{2}=-e^{2 \nu} d t^{2}+e^{2 \alpha}\left(d r^{2}+r^{2} d \theta^{2}\right)
+e^{2 \beta} r^{2} \sin ^{2} \theta(d \phi-\omega d t)^{2},
\end{split}
\end{equation}
where $\nu,~\alpha,~\beta, ~\omega$ are  functions of $r$ and $\theta$. Therefore, the model of the rotating neutron star will be the axisymmetric solution of the Einstein field equation, and  the material inside the neutron star can be still assumed  as perfect fluid. In order to solve the field equation for the axisymmetric solution, we also need the equation of hydrostatic equilibrium \cite{Komatsu1989,Nozawa1998}
\begin{footnotesize}
\begin{equation}
\nabla p+(\varepsilon+p)\left[\nabla \nu+\frac{1}{1-\textsl{v}^{2}}\left(-\textsl{v} \nabla \textsl{v}+\textsl{v}^{2} \frac{\nabla \Omega}{\Omega-\omega}\right)\right]=0,
\end{equation}
\end{footnotesize}
where $\nabla$  is the 3-dimensional derivative operator with spherical polar coordinates $r,~\theta,~\phi$, and $\textsl{v}$ is the  3-velocity. To integrate the above formulas, we need $u^{t} u_{\phi}=j$, where $u^{t},~u_{\phi}$ are the four-velocity, and $j$ is the function of angular velocity $\Omega$.

 The angular momentum is given by \cite{Komatsu1989,Nozawa1998}
\begin{equation}
J=2 \pi \int d r \int d \theta r^{3} \sin ^{2} \theta \frac{(\varepsilon+p) \textsl{v}}{1-\textsl{v}^{2}} e^{2 \alpha+2 \beta},
\end{equation}
and then we can get the inertia of momentum through $I=J / \Omega$.

The redshift in the rotating case is a little bit more complicated than that in the static case, as the  redshift in the rotating case includes two parts: the gravitational redshift and the doppler redshift generated by the rotation. Therefore,  if we observe the redshift from different directions, the results will be different.
To simplify the discussion, here we only consider the polar redshift, which is given by \cite{Cook1994}
\begin{equation}
Z_{p}=e^{-\left(\gamma+\rho\right) / 2}-1
\end{equation}
where $\gamma=\beta+\nu$ and $\rho=\nu-\beta$.

In the rotating case, the definition of gravitational mass and proper mass can be found in Ref. \cite{Cook1994}, and the definition of binding energy is similar to the static case.

In this work, we use the RNS \cite{49} program to perform the numerical calculation.

\section{Equations of state}\label{sec3}

As we know, there are still great difficulties to extrapolate the current knowledge of the EOS at the normal nuclear density into the super high density   in neutron star core. At present, there are thousands of EOS models used  to describe the neutron star matter. Different EOS models often incur significant divergence
at supra-saturation densities. In fact, this is the exact   reason why we are trying to seek  for the universal relations.
In order to make the adopted EOSs representative and satisfy the known constraints,  we  selected 11 EOSs, which can support a maximum mass of neutron star higher than $2.01~ M_{\odot}$ and are described under the predictions of microscopic nuclear many-body theories for normal neutron stars and  hybrid stars.
The 11 microscopic EOSs are as
follows: ALF2 by Alford et al. \cite{34}; ENG by L. Engvik et al. \cite{35}; MPA1 of Muther et al. \cite{36}; WWF1 and WWF2F of RBWiringa et al. \cite{37}; SLY of F. Douchin and P. Haensel \cite{38}; APR3 and APR4 of Akmal and VR Pandharipande \cite{39}; QMFL40, QMFL60, and QMFL80 of Zhu et al. \cite{40}. The details of these EOSs can be referred to in the corresponding literatures.

\begin{figure}[!htb]
	\centering
	\includegraphics[width=0.6\textwidth]{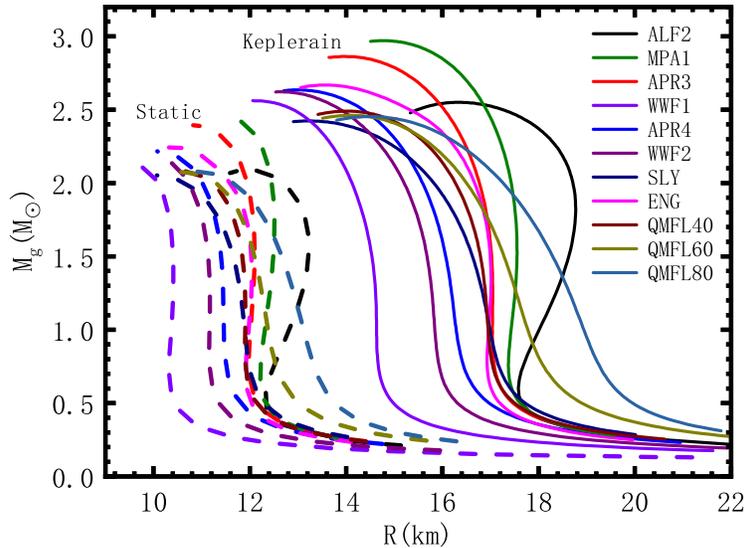}
	\caption{ Mass-radius relation for static and  Keplerian rotating neutron stars. }
\label{Fig.1}
\end{figure}

The mass-radius relations for the static and maximally rotating (that is, rotating at Keplerian frequency) neutron stars are presented in Fig. \ref{Fig.1}. The wider distribution area of the mass-radius relation indicates that this set of EOSs is   representative.

\section{Result and discussion}\label{sec4}

In this section, we   present the results in three cases  (that is, the  static (non-rotating) case, the general rotating case and Keplerian  rotating case). In all cases, the fitting formula takes the following form
\begin{equation}
\label{fit}
Y^{fit}=\sum_{n}A_{n}x^{n},
\end{equation}
where $x, Y^{fit}$ correspond to the horizontal and vertical coordinates of the universal relation figures. The coefficient $A_{n}$ of all the universal relations obtained in this work are presented in the Table \ref{tab1}.

\begin{table*}[!htb]
	\centering
	\caption{The fitting parameter}
	\label{tab1}
	\begin{ruledtabular}
		\begin{tabular}{ccccccc}
			Fig.&$A_0$  &$A_1$  &$A_2$  &$A_3$  &$A_4$   \\ \hline
			\ref{Fig.2}&3.646  &-10.54  &16.75  &-13.42  &4.865\\
            \ref{Fig.2b}&-0.0034  &0.2431  &0.6892  &-0.9030  &0.9221\\
			\ref{Fig.3}&0.0125  &0.5994\\
            \ref{Fig.4a}&3.013  &-1.515  &0.6256  &-0.1361  &0.0107\\
			\ref{Fig.5}&3.057  &-1.690  &0.7326  &-0.1597  &0.0127\\
			\ref{Fig.6}&-0.1855  &6.894  &-1.794\\
			\ref{Fig.7}&0.0017  &1.0572  &-1.218\\
		\end{tabular}
	\end{ruledtabular}
\end{table*}

\subsection{Universality in static cases}

   As we know, all of the observed neutron stars (pulsars) are rotating. Normally, a neutron star deformed by rapid rotation will deviate from spherical symmetry. But when the spin frequency is lower than 200 Hz,  a  spherically symmetrical static  star model can be looked on as  an accurate representation of the stars \cite{43,44}.  In fact, for some quantities in the universal relation, their definitions are based on rotation, such as the moment of inertia ($I=J/\Omega$). But in many cases a static model is adopted to perform the numerical  calculation \cite{9,12,13}.

    Here we will present a group of universal relations involving the gravitational redshift, moment of inertia and gravitational binding energy for static cases, where some of the quantities  are scaled by the stellar mass or the radius.

\begin{figure}[!htb]
	\centering
	\includegraphics[width=0.6\textwidth]{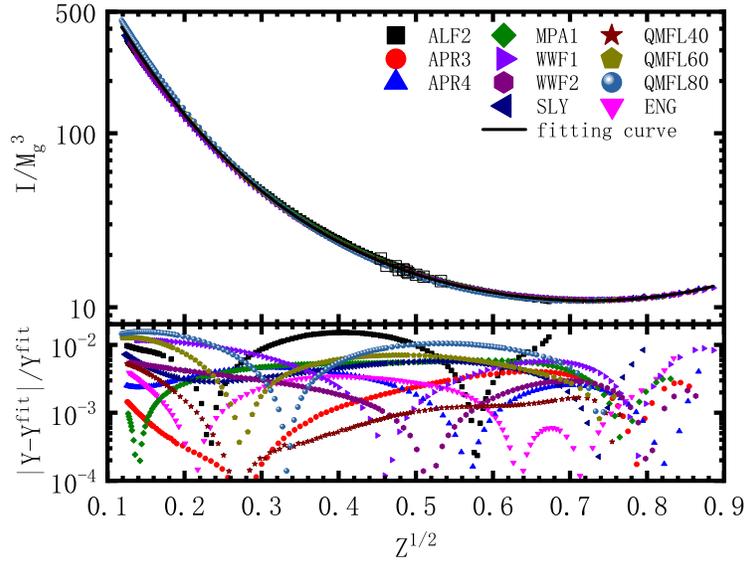}
	\caption{ (Upper panel) Universal relation between  the dimensionless moment of inertia and  the  square root of   redshift, where the  moment of inertia is scaled by the gravitational mass  as $I/M_g^{3}$,  and the hollow boxes mark the points with mass of $1.4~M_{\odot}$. (Lower panel) The relative error
between the fitting curve and the numerical results, where $Y=\log_{10} (I/M_g^{3})$. }
\label{Fig.2}
\end{figure}

\begin{figure}[!htb]
	\centering
	\includegraphics[width=0.6\textwidth]{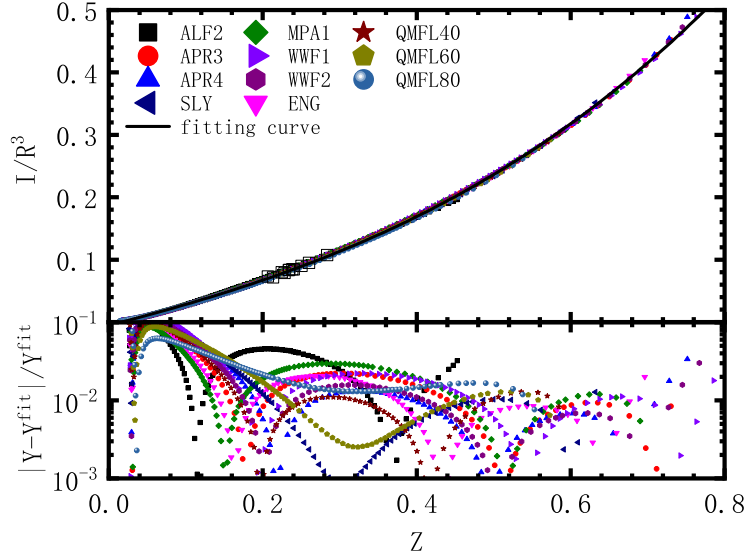}
	\caption{ (Upper panel) Universal relation between  the dimensionless moment of inertia and  the  redshift, where the  moment of inertia is scaled by the radius as $I/R^{3}$,  and the hollow boxes mark the points with mass of $1.4~M_{\odot}$. (Lower panel) The relative error
between the fitting curve and the numerical results, where $Y=I/R^{3}$.}
 \label{Fig.2b}
 \end{figure}

\begin{figure}[!htb]
	\centering
	\includegraphics[width=0.6\textwidth]{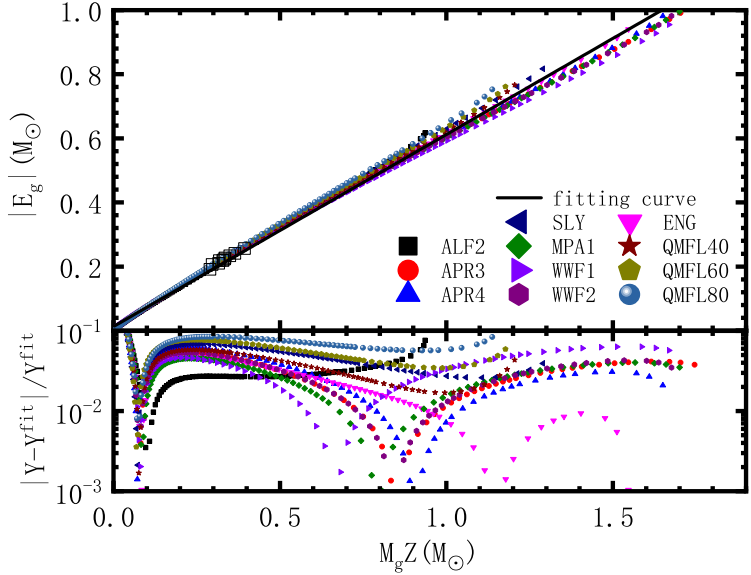}
	\caption{
 (Upper panel) Universal relation between  the gravitational binding energy and the gravitational redshift, where the gravitational redshift is scaled by the gravitational mass  as $M_{g}Z$, and the hollow boxes mark the points with mass of $1.4~M_{\odot}$. (Lower panel) The relative error between the fitting curve and the numerical results, where $Y=|E_{g}|$.}
	\label{Fig.3}
\end{figure}

 Considering  the  gravitational  redshift is defined directly through the compactness ($C=M_{g}/R$, see Eq. \ref{GZ}) and the dimensionless moment of inertia ($\bar{I}=I/M_g^{3}$) has a universal relation with the compactness \cite{9,Chan2016}, it is natural to expect that there will be a  universal relation  between  these two quantities. As shown  in Fig. \ref{Fig.2},  there is indeed a universal relation between  the dimensionless moment of inertia ($\bar{I}=I/M_g^{3}$) and  the  square root of  gravitational  redshift ($Z^{1/2}$), where the hollow boxes present the result of canonical neutron stars with mass of $1.4~M_{\odot}$. As presented in the lower panel of this figure,  most of the  relative errors are lower than 1\%.
 It's worth noting  that the relative error will have about an order of magnitude difference if the vertical coordinates  adopt  the linear or the logarithmic coordinates, as shown in Fig. 2 of this work,  Fig. 2 of Ref. \cite{Chan2016}  and Fig. 15 of Ref. \cite{9}.
 In this universal relation, there are three quantities (moment of inertia, gravitational mass, gravitational redshift) involved in this universal relation, which means that if any two of these quantities are observed for a neutron star, the other one can be estimated through this universal relation. As pointed out in  literatures, the moment of inertia can be observed by accurate pulsar timing in a binary system \cite{45,46}. In the near future, if both the mass and the moment of inertia of a neutron star are observed simultaneously,  this universal relation provides a way to estimate the gravitational redshift.

The radius of neutron star is an important but difficult to be accurately observed global property. There are many approaches to estimate or constrain the radius, such as estimation  from GW170817 (please refer to a summary in Ref. \cite{Baiotti2019}) or measurement based on x-ray, optical and radio observations \cite{Watts2016}. Universality provides another potential way to estimate the radius. Through scaling the  moment of inertia by the radius as $I/R^{3}$, we obtained a universal relation between  the dimensionless moment of inertia and  the  gravitational redshift, as shown in  Fig. \ref{Fig.2b}. Obviously, as long as both of the moment of inertia and the gravitational redshift are observed  or derived  through universal relation,  the radius can be constrained  through this universal relation.

In Fig. \ref{Fig.3}, we present the universal relation between  the gravitational binding energy ($E_{g}$) and the mass scaled gravitational redshift ($M_{g}Z$). Clearly, this universal relation can be well approximated by a linear fitting, as the fitting coefficient $A_{n}$ shown in  Table \ref{tab1}. This universal relation can be used to estimate the gravitational binding energy if the stellar mass and the gravitational redshift of a neutron star are obtained.

\subsection{Universality in general rotating case}

The currently known fastest spin frequency (716 Hz) of  pulsar PSR J1748-2446ad \cite{47} is  much lower than the Keplerian frequency supported by a canonical neutron star model \cite{Cook1994,Wen2011}. Therefore, to investigate the universality of neutron stars with considerable rotation effect, we set a   general rotational  frequency range  as  [200, 900] Hz, and perform the numerical calculation  at 100 Hz intervals from 200 Hz to 900 Hz.

\begin{figure}[!htb]
	\centering
	\includegraphics[width=0.6\textwidth]{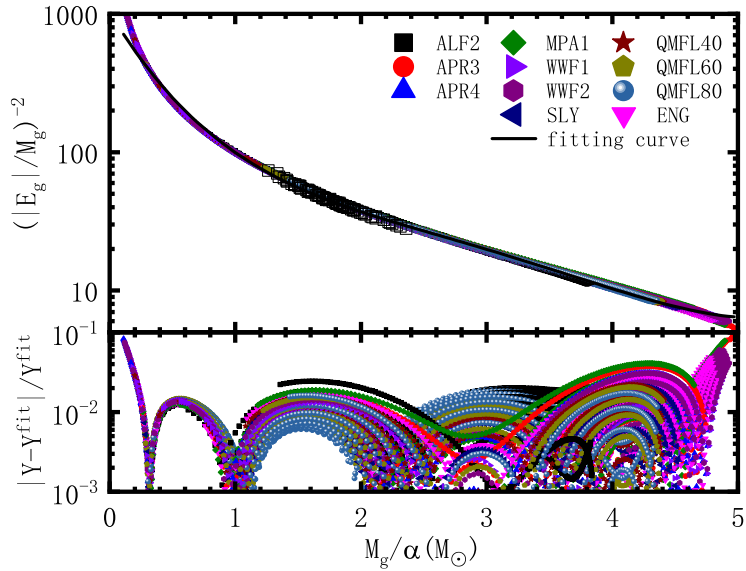}
	\caption{
 (Upper panel) Universal relation between $(|E_{g}|/M_{g})^{-2}$ and $M_{g}/\alpha$ for the general rotating case, where $\alpha=J/M_{g}^{2}$, and  the hollow boxes mark the points with mass of $1.4~M_{\odot}$. (Lower panel) The relative error between the fitting curve and the numerical results, where $Y=\log_{10} (|E_{g}|/M_{g})^{-2}$.}
\label{Fig.4a}
\end{figure}

\begin{figure}[!htb]
	\centering
	\includegraphics[width=0.6\textwidth]{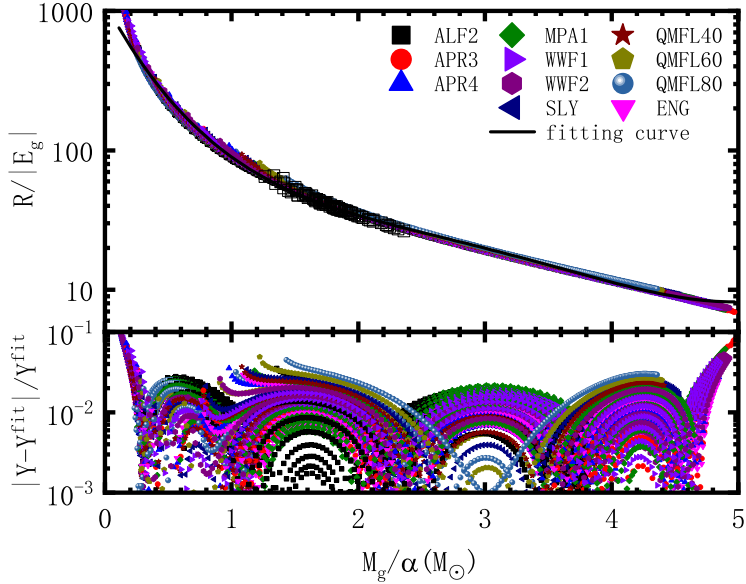}
	\caption{
 (Upper panel) Universal relation between $R/|E_{g}|$ and $M_{g}/\alpha$  for the general rotating case, where  the hollow boxes mark the points with mass of $1.4~M_{\odot}$. (Lower panel) The relative error between the fitting curve and the numerical results, where $Y=\log_{10}( R/|E_{g}|)$.}
 	\label{Fig.5}
\end{figure}

Firstly, we find  that there exists a universal relation between these two combinatorial quantities:  $(|E_{g}|/M_{g})^{-2}$ and $M_{g}/\alpha$ (where $\alpha=J/M_{g}^{2}$) for the neutron stars with rotational frequency in   [200, 900] Hz, as shown in Fig. \ref{Fig.4a}. Usually,  a neutron star is identified by the observation of pulse (normally being looked on as the spin frequency of pulsar/neutron star), thus the spin frequencies  of   observed neutron stars are default known. Therefore, if the mass ($M_{g}$) and moment of inertia ($I$) of a neutron star with fast spin frequency is observed, then we can estimate its gravitational binding energy through this universal relation, noting that $J=I\Omega$. It is worth pointing out that to estimate the gravitational binding energy is not only for the understanding of this  unobservable quantity itself, but also for the further constraint on the radius of neutron star, which  will  be presented next.

 As presented in  Fig. \ref{Fig.5}, we obtain another universal relation based on the $M_{g}/\alpha$ , that is, the relation between   $R/|E_{g}|$ and $M_{g}/\alpha$. Based on the discussion of  Fig. \ref{Fig.4a}, if the  gravitational binding energy $E_{g}$ is estimated by the way given above, then we can further estimate the radius through the universal relation of  Fig. \ref{Fig.5} by the similar way we used to estimate the $E_{g}$ .

\subsection{Universality in Keplerian rotating case}

 The ultimate  properties of neutron stars, such as the maximum stellar mass and the shortest spin period,  are often concerned in the study of compact stars. These properties are related to the mass-shedding rotation, that is, the Keplerian rotation, which could be determined  by setting a test particle rotating in a stable circular orbit at the equator, where  the centrifugal force are just balanced by the gravitational force. From the observational point of view, the Keplerian rotation, which is closely related to the kilohertz quasi-periodic oscillations, has been well observed in low-mass X-ray binaries (LMXBs) \cite{Luk2018}.

 Similarly, due to the uncertainty of the EOS,  finding  the universality between the properties of neutron stars in Keplerian rotating case and in the static case provides  a practical way to understand these ultimate properties \cite{26}.

\begin{figure}[!htb]
	\centering
	\includegraphics[width=0.6\textwidth]{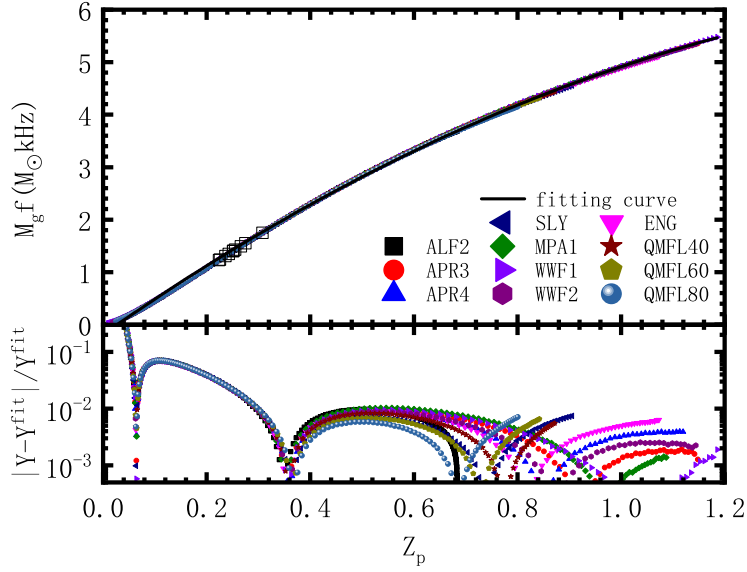}
	\caption{
(Upper panel) Universal relation between $M_{g}f$ and $Z_p$  for the Keplerian rotating case, where $f=\Omega/(2\pi)$ is the spin frequency, $Z_p$ is the polar redshift, and  the hollow boxes mark the points with mass of $1.4~M_{\odot}$. (Lower panel) The relative error between the fitting curve and the numerical results, where $Y=M_{g}f$.}
	\label{Fig.6}
\end{figure}

\begin{figure}[!htb]
	\centering
	\includegraphics[width=0.6\textwidth]{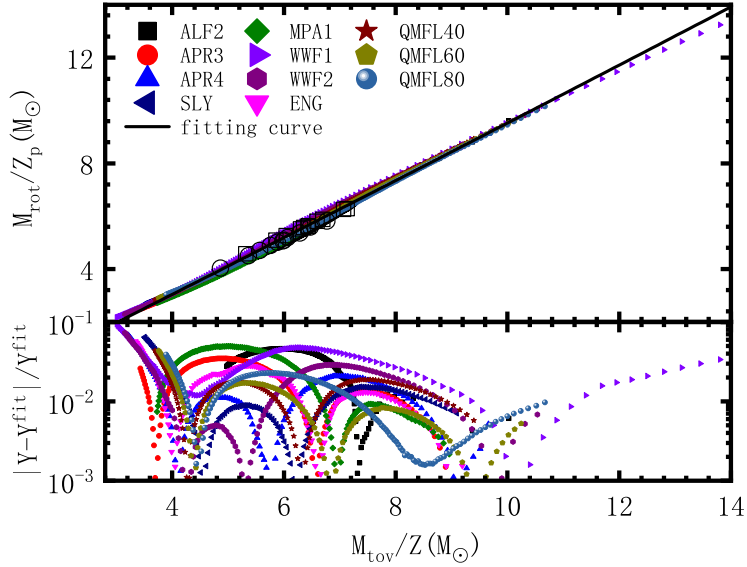}
	\caption{
(Upper panel) Universal relation between $M_{\textrm{rot}}/Z_{\textrm{p}}$ and $M_{\textrm{tov}}/Z$,  where $M_{\textrm{tov}}$  and $Z$ are the mass and redshift of the static case, $M_{\textrm{rot}}$ and $Z_{\textrm{p}}$ are the mass and polar redshift of
 Keplerian rotating case, and  the hollow circles mark the points of static stars with mass of $1.4~M_{\odot}$ while the hollow boxes mark the points of  Keplerian rotating stars with mass of $1.4~M_{\odot}$. (Lower panel) The relative error between the fitting curve and the numerical results, where $Y=M_{\textrm{rot}}/Z_{\textrm{p}}$.}
	\label{Fig.7}
\end{figure}

For the neutron stars rotating at the Keplerian frequency, it is shown that there is a perfect universal relation between the mass-scaled spin frequency $M_{g}f$ and the polar redshift $Z_{\textrm{p}}$, as shown in Fig. \ref{Fig.6}.
For the EOSs  adopted here, the  Keplerian rotating canonical neutron stars ($1.4~M_{\odot}$)  have polar redshift within $\sim [0.2,~0.3]$.
If we take the known fastest spin frequency 716 Hz as the Keplerian frequency of a canonical neutron star, then its polar redshift will take a value around $0.2$. For a  neutron star with  mass of $2.0~M_{\odot}$ and with  Keplerian frequency of 2 kHz, its polar redshift will be about 0.8.

 As mentioned above, it is interesting to probe the universality between the  quantities of static and Keplerian rotating cases, which could provide an EOS-independent connection between the non-rotating and the fastest rotating neutron stars \cite{26}. In  Fig. \ref{Fig.7}, we present the universal relation between $M_{\textrm{rot}}/Z_{\textrm{p}}$ and $M_{\textrm{tov}}/Z$,  where $M_{\textrm{tov}}$  and $Z$ are the mass and gravitational redshift of the static stars, $M_{\textrm{rot}}$ and $Z_{\textrm{p}}$ are the mass and polar redshift of Keplerian rotating stars. As shown in this figure, the fitting curve can be approximated as a straight line. It is worth mentioning that  the non-rotating stars and the fastest rotating  stars are connected by the same central density. As shown in Table \ref{tab1}, the fitting parameter $A_{1}=1.0572$, thus the fitting line  can be approximated as $M_{\textrm{tov}}/Z \approx M_{\textrm{rot}}/Z_{\textrm{p}}$. As we know, at the same central density, a rotating star has a higher mass than that of the static one. It is natural to get a conclusion that at the same central density,   the polar redshift (essentially being also the gravitational redshift) of the Keplerian rotating star is higher than the gravitational redshift of the static star.


\section{Conclusion}\label{sec5}

Based on 11 equations of state (EOSs) from the predictions of microscopic nuclear many-body theories for normal and hybrid neutron stars, a group of new universal relations of gravitational redshift, moment of inertia (or angular momentum) and gravitational binding energy are proposed, where these quantities are  usually scaled by the mass or the radius. These universal relations can be  summed up in the following three cases.

\begin{enumerate}[(1)]
	\item
	In the case of non-rotation (static), universal relations between  the dimensionless moment of inertia and  the  square root of  gravitational  redshift,  between  the dimensionless moment of inertia and  the  gravitational redshift and  between  the gravitational binding energy  and the mass-scaled gravitational redshift are obtained. These universal relations provide us a way to estimate the redshift, the radius and gravitational binding energy if the mass and moment of inertia of a neutron star are accurately observed.
	\item
	In the case of general rotation, we find  the universality between $(|E_{g}|/M_{g})^{-2}$ and $M_{g}/\alpha$ and the universality between   $R/|E_{g}|$ and $M_{g}/\alpha$. These universal relations can be used to estimate the gravitational binding energy and the radius of fast rotating neutron star.
	\item
	In the case of Keplerian rotation, the universality between $M_{g}f$ and $Z_{\textrm{p}}$ and universality between $M_{\textrm{rot}}/Z_{\textrm{p}}$ and $M_{\textrm{tov}}/Z$ are obtained. These universal relations provide a way to constrain the polar redshift and build up  a  connection between the properties of non-rotating and   fastest rotating neutron stars.

\end{enumerate}

These new universal relations provide a potential way to constrain or estimate the unobservable  or hard to be observed properties of neutron stars.
It is also an interesting topic to probe the universality related to the parameters of EOS or the internal parameter of neutron star (such as the
central density). We hope to return to this issue in the future.

\section{Acknowledgements}

We thank Zhao-Qing Feng and Xiang-Dong Zhang for helpful discussions. This work is supported by NSFC (Grants No. 11975101 and No. 11722546), Guangdong Natural Science Foundation (Grant No. 2020A1515010820) and the talent program of South China University of Technology (Grant No. K5180470). This project has made use of NASA's Astrophysics Data System.


%

\end{document}